# A mixed-signal ASIC for the readout of Gas Electron Multiplier detectors
## Design review and characterization results

Fabio Cossio
National Institute of Nuclear Physics
Turin Polytechnic Ph.D. student
Turin, Italy

*Abstract*—TIGER (Turin Integrated Gem Electronics for Readout) is a mixed-mode front-end ASIC developed to readout the new inner tracking detector of the BESIII experiment, carried out at BEPCII in Beijing. The detector is planned to be installed during the 2018 upgrade and features an innovative three-layer triple-CGEM (Cylindrical Gas Electron Multiplier) with analog readout. The ASIC comprises 64 channels, each of which features a dual-branch architecture to extract and digitize the timestamp and charge of the input signal. The time-of-arrival is provided by a set of low-power TDCs, based on analog interpolation techniques, while the charge measurement is obtained either from the time-over-threshold information or the 10-bit digitization of the signal peak amplitude. Fabricated in a 110 nm CMOS technology, the ASIC has been designed to operate with an input capacitance of about 100 pF, an input dynamic range between 1 and 50 fC, a power consumption of 10 mW/channel and a sustained event rate of 60 kHz/channel.

## I. Introduction

The BESIII (BEijing Spectrometer III) experiment studies the events produced in the annihilation of electrons and positrons colliding in the BEPCII (Beijing Electro-Positron Collider II) accelerator, at the IHEP (Institute of High Energy Physics) laboratory in Beijing. To overcome the aging effect, due to radiation damage, of the inner part of the MDC (Multilayer Drift Chamber), an upgrade based on CGEM technology has been proposed. The project features three independent tracking layers, each made of a cylindrical triple-GEM detector. A GEM (Gas Electron Multiplier) is made by a 50 μm Kapton foil, copper clad on each side and with a high density of holes. Inside the holes, an electric field of 100 kV/cm multiplies the number of the electrons produced by a charged particle crossing the detector by a factor up to a few thousand [1]. Stacking together three GEMs allows to reach high gains while minimizing the discharge probability. Each layer of the BESIII CGEM detector is composed by five concentric cylindrical electrodes: the cathode, three GEM foils, and the readout anode. A schematic of a typical triple CGEM is depicted in Fig. 1. The readout anode of each layer is segmented with 650 μm pitch XV patterned strips with different stereo angles depending on the layer geometry [2].

To achieve the required spatial resolution of the order of 100 μm both binary and analog strip readout methods have been investigated. Due to the relatively large BESIII magnetic field, the analog readout is mandatory since it offers the best trade-off between the spatial resolution and the number of channels (about 10.000). In fact, the strip collected charge information allows the reconstruction of the charge centroid, thus improving the resolution compared to the one achievable with digital information, which is limited to pitch/$\sqrt{12}$ [3]. Furthermore, the time information allows to reconstruct the particles tracks and operate the CGEM detector as a "μ-TPC" (micro Time Projection Chamber) thus improving the spatial resolution of angled tracks. This solution, on the other hand, requires a dedicated analog readout front-end ASIC that will be discussed in this paper. In the following section, some aspects of the design of the ASIC are reviewed, starting from its requirements and focusing on time and charge measurement principle of operation. Then the test setup used to characterize the chip prototype is presented, as well as the first results from the silicon tests carried out to validate the ASIC.

## II. ASIC Architecture

TIGER is a mixed-signal ASIC for the readout of the CGEM detector. It consists of 64 channels, references and bias generators, a digital global controller and an internal test pulse calibration circuitry. It has been designed in a 110 nm CMOS technology and inherits an SEU-upgraded version of the digital backend of the TOFPET2, an ASIC developed for the readout of signals from Silicon Photomultipliers (SiPM) in PET (Positron Emission Tomography) applications [4-5].

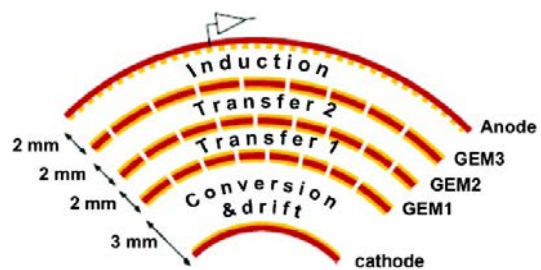

Fig. 1. Schematic of a triple cylindrical GEM detector.

## A. Requirements

The charge centroid method requires the ASIC to provide not only the fired-strip single-bit information but also the measurement of the charge collected by each strip, with a typical input signal expected to be in the range between 1 and 50 fC. Moreover, a time resolution better than 10 ns is required to time-tag each hit and then operate the CGEM as a µTPC. The CGEM strips have a quite large pitch (650 µm) and can be up to 82 cm long: this leads to a parasitic capacitance at the input of each channel of about 100 pF. Furthermore, the space available in the inner part of the BESIII spectrometer is quite limited, thus introducing a 10mW/channel power dissipation constraint.

## B. Channel architecture

The architecture of one channel is shown in Fig. 2. It can be divided in two parts: the front-end receives the CGEM signals that are amplified and shaped, while the back-end performs signal digitization, data storing and transmission. The front-end is composed of a charge sensitive amplifier, which generates two replicas of the amplified signal that are sent to two shaping amplifiers: the Time-branch shaper generates a signal with a steep slope (50 ns peaking time), optimized for timing measurements with a leading-edge threshold, while the Energy-branch shaper provides a more flattened signal (160 ns peaking time), allowing for better signal integration and ENC (Equivalent Noise Charge) optimization. The outputs of the two shapers are fed to two discriminators with separate and programmable thresholds, set by 6-bit DACs with a configurable range and LSB, enabling for different operating modes. The discriminators outputs are sent to the channel controller, a digital logic block working at 160MHz clock frequency, that generates the trigger signals that control the channel operations. Each channel includes two time-to-digital converters (TDC) and a sample-and-hold circuit (Peak Detector). The time of the event is obtained from a 16-bit coarse timestamp (coarse time) and a 10-bit fine measurement (fine time). The coarse time is provided by a global clock counter, with a coarse time resolution of 6.25 ns for a master clock frequency of 160 MHz. The fine time information indicates the phase between the asynchronous trigger, generated by the discriminator, and the next clock edge and it is measured with an analog low-power TDC, composed of a set of 4 time-to-analog converters (TAC) and a Wilkinson ADC [6-7]. The time-to-analog conversion principle of operation is depicted in Fig. 3. For each trigger, a constant current discharges a capacitor until the next known clock phase is reached. After that, the voltage is transferred into a 4x bigger capacitor and digitized with a Wilkinson ADC by recharging the capacitor with a 32x smaller current. This results in a time interpolation factor of 128, delivering a 50 ps time binning at 160 MHz clock frequency. When the capacitor voltage reaches a programmable reference voltage, a latched comparator stops the recharge. A multi-buffer scheme allows to de-randomize the input event rate and lessen the issue of the inherently high conversion time of this approach.

A sample-and-hold circuit (S/H) samples the output signal of the E-branch shaper within a time window, generated by the digital logic and configurable by the user. This signal is then converted into a 10-bit value (Efine) with the same Wilkinson ADC used for the TDC operation, therefore the same quad-buffer approach used for the TACs is implemented also for the S/H circuit.

Thanks to its dual-threshold architecture and some configurable logics, the ASIC can operate in different modes. In S/H mode, the leading edge trigger on the T-branch generates the event timestamp while the charge measurement is provided by the S/H circuit of the E-branch. In ToT mode both branches provide a time measurement and the two timestamps allow to compute the time-over-threshold, which is dependent on the signal amplitude.

A test pulse (TP) signal can be sent either to the front-end to test the whole readout chain or used directly by the channel controller to test and calibrate the TDCs. The test pulse sent to the front-end is an analog signal generated by an internal calibration circuit and its amplitude is user-configurable, allowing to test the ASIC with different input charge values.

## C. Global controller operation

The readout scheme employs a data-push architecture: each input signal above the threshold is digitized and sent to the global controller that manages the off-chip data transmission. Data are serialized and transferred to an FPGA board through 2 LVDS links, using 8B/10B encoding and TX training for synchronization with the FPGA. The channels and global registers are configured through a 10 MHz SPI like interface. The digital logic of the ASIC has been protected against Single Event Upset (SEU) by using Hamming encoding and error correction techniques.

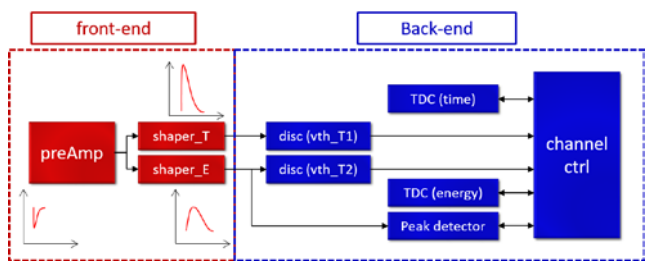

Fig. 2. TIGER channel architecture.

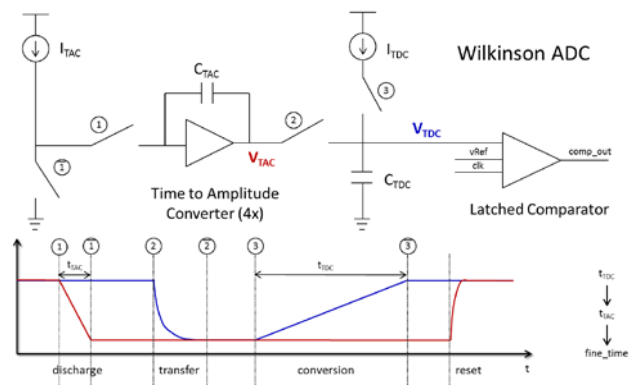

Fig. 3. TDC principle of operation.

## III. TEST SETUP

Fig. 4 shows the test bench setup used to characterize the TIGER prototype: the setup includes an FPGA board, power supplies, a low-jitter reference clock source and a pulse generator.

The ASIC test board was designed, fabricated and assembled by the Turin INFN electronics group and is equipped with five trimmers to correctly set the ASIC biasing, 2 SMA LVDS inputs to feed the clock to the chip and 8 SMA LVDS outputs used for debug. The test board also provides direct outputs from the shaper and threshold of the channel 63 T-branch. These signals are seen at the oscilloscope to infer some useful information during the debug-test phase, e.g. the input charge dynamic range, the front-end gain, the LSB of the DAC that sets the threshold. Separated DC power supplies are used for the analog and the digital domains, using an Agilent E3631A 80W Triple Output Power Supply module.

The FPGA board is a Xilinx Virtex 6 Evaluation board (ML605 Evaluation kit) and it is connected to the test board with an FMC connector (up to two test boards can be connected to the same FPGA board). It handles the ASIC configuration and data reception and generates a test pulse with adjustable timing which is delivered to the ASIC for testing purposes. An Ethernet cable connects the FPGA board to a PC where a LabVIEW program manages the data acquisition. The 160 MHz clock can be synthesized either by the FPGA board or with a low-jitter Stanford Research CG635 Clock Generator (1 µHz to 2.05 GHz square wave generator). An Agilent 81133A Single Channel 3.35 GHz Pulse Pattern Generator was used to inject a known charge at the input of channel 63 in order to calibrate the internal test pulse (the pulse generator was locked to the clock generator using the 10 MHz output of the CG635 module).

## IV. RESULTS

### A. TDC calibration and time resolution

The TDCs of the two branches are scanned over dynamic range by sweeping the TP phase along one clock cycle. A LUT (look-up table) with gain and offset correction for all channels is generated. Fig. 5 shows that, after the calibration, the TDCs quantization error is lower than 50 ps r.m.s. for all the channels, such that its contribution to the intrinsic time resolution becomes negligible.

### B. S/H linearity

In order to test the linearity of the S/H circuit, signals ranging from 6 fC to 50 fC have been fed to the input of channel 63, using the Agilent pulse generator. Fig. 6 shows the residual non-linearity, defined as the relative difference between the S/H measured value and the linear fit. The linearity is very good up to 40 fC (less than 0.2%). With the information extracted from the external pulse generator, the internal TP has been calibrated and then sent to the other channels, which showed a very similar response.

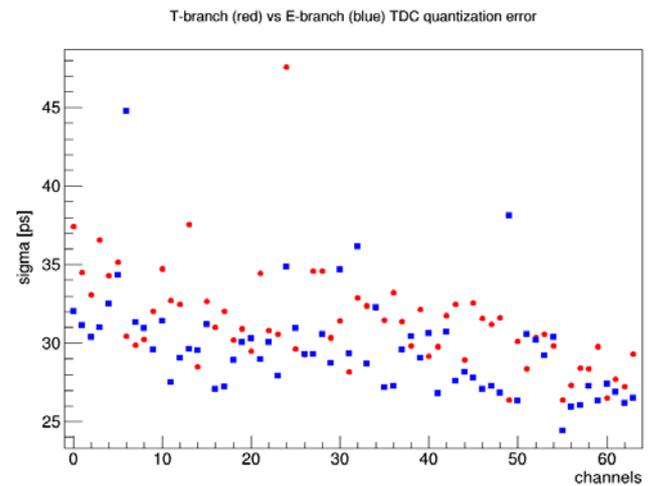

Fig. 5. T-branch (red) and E-branch (blue) TDCs quantization error.

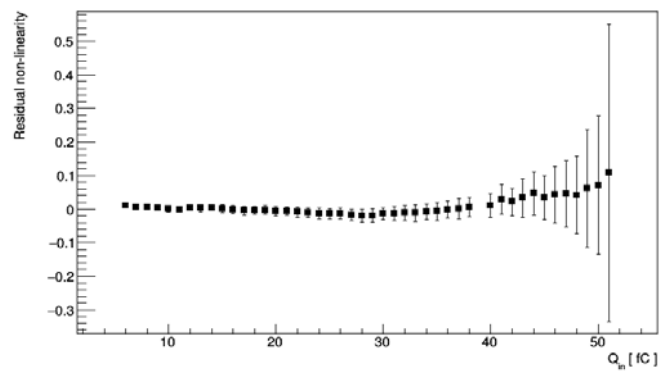

Fig. 6. S/H residual non-linearity (external TP, channel 63).

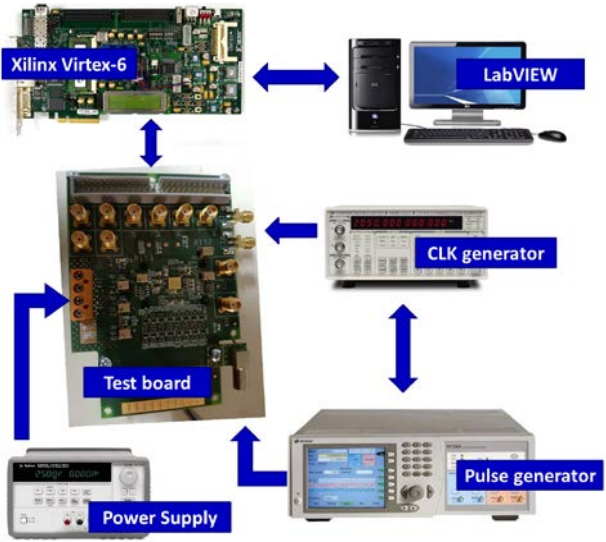

Fig. 4. Test bench setup.

## C. Threshold equalization

The results of a threshold scan for the 64 channels are shown in Fig. 7: the peaks indicate the position of the baseline and show the spreading of the baseline among the channels. As a consequence, setting the same nominal value, i.e. the same 6-bit DAC code, for all the channels will result in different effective thresholds. A threshold equalization mitigates this effect and works as follows: for each channel, the difference between its baseline and the channels mean baseline is evaluated. This difference is then used as a reference to generate a LUT that separately sets the threshold of each channel a fixed number of LSB above the baseline. Fig. 8 illustrates the ToT measurement for the 64 channels before and after threshold equalization, showing a more uniform distribution after the equalization. The non-linearity of the ToT curve comes from the shapers intrinsic non-linearity and can be corrected offline by means of a calibration run.

## D. Noise measurement

The threshold scan allows also to evaluate the front-end channel noise. For this measurement, an input signal of fixed amplitude is sent to the front-end. The S-curve produced by the threshold scan is fitted with a sigmoid function and its sigma value provides the noise at the output of the shaper. This method has been iterated for different input capacitances. The noise measured at the output of the T-branch shaper with a 100 pF input capacitance is about 4 mV, which is slightly bigger than the value expected from simulations but still adequate for our application. Nevertheless, in order improve the ASIC noise performance, PSRR (power supply rejection ratio), interference and grounding conditions are currently under study.

## V. CONCLUSIONS

A 64 channels ASIC, named TIGER, has been developed to readout the signals from the BESIII CGEM detector. The ASIC has been electrically characterized and all the specified requirements are met within the limited power budget of 10 mW per channel. In the next months the ASIC will be tested with the CGEM prototype for the final validation. After that, the chips will be produced in a dedicated run for the 2018 installation.

## ACKNOWLEDGMENT

This research activity has been performed within the BESIIICGEM Project, funded by European Commission in the call H2020-MSCA-RISE-2014.

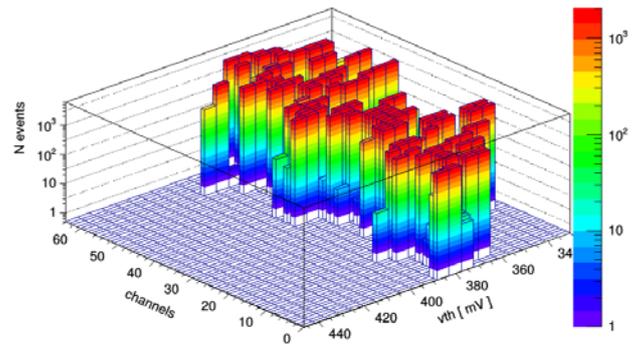

Fig. 7. Baseline spreading among the 64 channels.

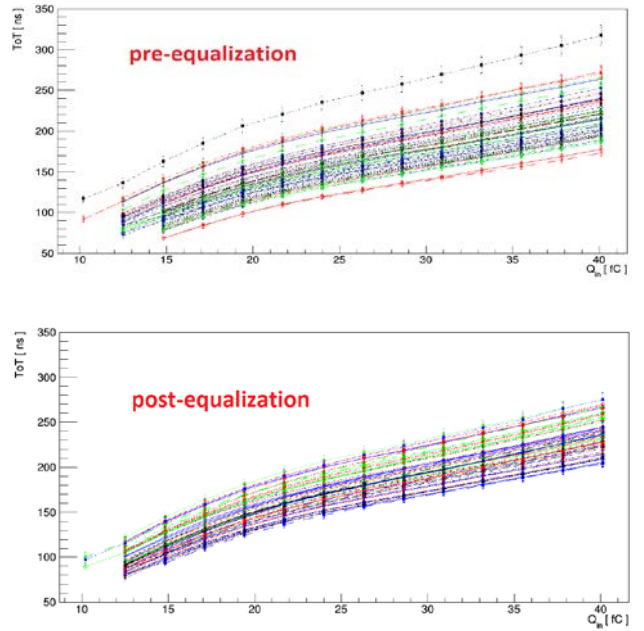

Fig. 8. Time-over-threshold as a function of the input charge (pre-equalization and post-equalization).